\documentclass[onecolumn, 12pt]{IEEEtran}
\usepackage{amsfonts,amssymb,amsmath,bm}
\usepackage{graphicx,subfigure}
\usepackage{cite,color}

\usepackage{geometry}
\newtheorem{lemma}{\textbf{Lemma}}[section]
\newtheorem{prop}{\textbf{Proposition}}[section]

\newcommand{\trace}{{\rm Tr}}
\newcommand{\rank}{{\rm Rank}}

\newcommand{\st}{{\rm s.t.}}

\newcommand{\bI}{\mathbf{I}}
\newcommand{\bH}{\mathbf{H}}
\newcommand{\bP}{\mathbf{P}}
\newcommand{\bG}{\mathbf{G}}

\newcommand{\bF}{\mathbf{F}}
\newcommand{\bQ}{\mathbf{Q}}

\newcommand{\bA}{\mathbf{A}}
\newcommand{\bB}{\mathbf{B}}
\newcommand{\bC}{\mathbf{C}}

\newcommand{\bN}{\mathbf{N}}
\newcommand{\bE}{\mathbf{E}}
\newcommand{\bU}{\mathbf{U}}
\newcommand{\bV}{\mathbf{V}}

\newcommand{\bX}{\mathbf{X}}

\newcommand{\bY}{\mathbf{Y}}

\newcommand{\bW}{\mathbf{W}}

\newcommand{\tV}{\mathbf{\tilde{\bV}}}
\newcommand{\bg}{\bar{g}}

\newcommand{\bZ}{\mathbf{Z}}

\newcommand{\bSigma}{\mathbf{\Sigma}}

\newcommand{\bTheta}{\mathbf{\Theta}}

\newcommand{\mP}{\mathcal{P}}
\newcommand{\bbS}{\mathbb{S}}
\newcommand{\mC}{\mathbb{C}}

\newcommand{\bx}{\bm{x}}

\newcommand{\by}{\bm{y}}

\newcommand{\hH}{\hat{\mathbf{H}}}

\newcommand{\mE}{\mathbb{E}}

\newcommand{\bu}{\bm{u}}
\newcommand{\bv}{\bm{v}}

\newcommand{\bs}{\bm{s}}

\newcommand{\bn}{\bm{n}}
\newcommand{\Cdom}{\mathbb{C}}

\newcommand{\cgauss}{\mathcal{CN}}
\ifCLASSINFOpdf
\else
\fi
\begin{document}
%
\title{Secure Beamforming For MIMO Broadcasting With Wireless Information And Power Transfer}
%
%
%

\author{Qingjiang~Shi,~
        Weiqiang Xu,~
        Jinsong Wu,
        Enbin~Song,
        Yaming Wang
        \thanks{This work is supported by the National Nature Science Foundation of
China under grant 61302076, 61374020, Key Project of Chinese Ministry of Education under grant 212066, Zhejiang Provincial Natural Science Foundation of China under grant LY12F02042, LQ12F01009, LQ13F010008, the Science Foundation of Zhejiang Sci-Tech University (ZSTU) under grant 1203805Y, and The State Key Laboratory of Integrated Services Networks, Xidian University under grant ISN14-08. }
        \thanks{Qingjiang Shi is with the School of Information and Science Technology, Zhejiang Sci-Tech University, Hangzhou, China, 310018. He is also with the The State Key Laboratory of Integrated Services Networks, Xidian University. (email: qing.j.shi@gmail.com).}
\thanks{Weiqiang Xu and Yaming Wang are both with the School of Information and Science Technology, Zhejiang Sci-Tech University, Hangzhou, China, 310018. }
\thanks{Enbin Song is with College of Mathematics, Sichuan University, Chendu, Sichuan 610064, China}
\thanks{Jinsong Wu is with Alcatel-Lucent Bell Labs, Shanghai, China}
}

\maketitle

\begin{abstract}
This paper considers a basic MIMO information-energy broadcast system, where a multi-antenna transmitter transmits information and energy simultaneously to a multi-antenna information receiver and a dual-functional multi-antenna energy receiver which is also capable of decoding information. Due to the open nature of wireless medium and the dual purpose of information and energy transmission, secure information transmission while ensuring efficient energy harvesting is a critical issue for such a broadcast system. Assuming that physical layer security techniques are adopted for secure transmission, we study beamforming design to maximize the achievable secrecy rate subject to a total power constraint and an energy harvesting constraint. First, based on semidefinite relaxation, we propose global optimal solutions to the secrecy rate maximization (SRM) problem in the single-stream case and a specific full-stream case. Then, we propose inexact block coordinate descent (IBCD) algorithm to tackle the SRM problem of general case with arbitrary number of streams. We proves that the IBCD algorithm can monotonically converge to a Karush-Kuhn-Tucker (KKT) solution to the SRM problem. Furthermore, we extend the IBCD algorithm to the joint beamforming and artificial noise design problem. Finally, simulations are performed to validate the effectiveness of the proposed beamforming algorithms.
\end{abstract}


\setlength{\baselineskip}{1.45\baselineskip}

%
\IEEEpeerreviewmaketitle

\section{Introduction}
Since battery technologies have not yet matched advances in hardware and software technologies, conventional battery-powered wireless systems suffer from short lifetime and require frequent recharging in order to maintain system operation. On the other hand, the rapid development of information and communication technologies demands a huge amount of energy consumption and thus notably contributes to global warming and environmental pollution. As a result, energy harvesting from the environment has recently drawn a lot of interest in both industria and academia\cite{Wu2012book}. Among the common environmental energy resources, radio signal is particular due to its conventional role of information carrier. Recent research results have shown that the functions of wireless communications and radio-based energy harvesting could be attained simultaneously, which have been termed as
(simultaneous) wireless information and power transfer (WIPT)\cite{ZhangArXiv,Xu2013,ZhouArxiv}.

WIPT has been studied for various communication systems in different context. For example, Zhang and Ho\cite{ZhangArXiv} considered a MIMO broadcast system made up of a transmitter, one information receiver (IR) and one energy receiver (ER), and investigated the relevant rate-energy region and optimal transmission schemes. Xu \emph{et. al.} \cite{Xu2013} investigated the optimal information/energy beamforming strategy to achieve the maximum harvested energy for multi-user MISO WIPT system with separated information/energy
receivers. Two practical receiver designs for WIPT were proposed in \cite{ZhangArXiv,ZhouArxiv}, namely, time switching (TS) and power splitting (PS). Based on the PS scheme, Shi \emph{et. al.} studied the optimal joint beamforming and power splitting (JBPS) to achieve the minimum transmission power of a multi-user MISO downlink system  subject to both signal-to-interference-plus-noise (SINR) constraints and energy harvesting constraints. The JBPS problem for MISO interference channel (IFC) was studied in \cite{Ottersten2013}. The works \cite{Shen2013,Clerckx13,Park2014} also considered interference channel with WIPT. Shen \emph{et. al.}\cite{Shen2013} studied transmitter design for sum-rate maximization with energy harvesting constraints in MISO IFC, while Park and Clerckx \cite{Clerckx13,Park2014} investigated transmission strategy for MIMO IFC with energy harvesting. Furthermore, WIPT has been investigated in other channel setups such as relay channels \cite{Chalise2012,Krikidis2012,Fouladgar2012,Nasir2013} and OFDM channels \cite{Huang2013,Zhou2013arXiv}.

The above research works have not considered security issues in WIPT. However, due to the open nature of wireless medium and the dual purpose of information and energy transmission, the wireless information in WIPT systems is more susceptible to eavesdropping. As one of the examples with security concerns, a dual-functional energy harvester, which is capable of both information decoding (ID) and energy harvesting (EH), may be a potential eavesdropper. Hence, security is an important issue in WIPT. Recently, physical layer security (PLS) technologies\cite{Debbah2009} have attracted a lot of attentions due to the potentials to ensure highly secure communications by exploiting some physical properties of wireless channels. Based on the PLS technologies, a very limited number of research works have considered secure communication in WIPT\cite{Liu20136,DWKNg2013,DWKNg201311,DWKNg201309,DWKNg201312,Zhu2014}. Liu \emph{et. al.} \cite{Liu20136} studied both the secrecy rate maximization problem and sum-harvested-energy maximization problem for a multi-user MISO WIPT system where one transmitter sends information and energy to one IR and multiple ERs. They proposed global optimal solutions to both problems by using semidefinite relaxation (SDR)\cite{Luo2010} and one-dimensional search. Considering conservative secrecy rate constraints, the works \cite{Zhu2014,DWKNg201312} investigated secure transmission in PS-based multi-user MISO WIPT systems and studied transceiver design to achieve the minimum transmission power. Ng. \emph{et. al.} \cite{DWKNg2013} extended the work \cite{DWKNg201312} and considered secure layered video transmission for PS-based downlink multicast systems using both information and energy beamforming. A chance constraint was introduced in \cite{DWKNg2013} to guarantee a minimum secrecy rate with a given probability while achieving the minimum transmission power. Furthermore, Ng. \emph{et. al.} \cite{DWKNg201311} advocated the dual use of both artificial noise\cite{Liao2011} and energy signals to provide both secure communication and efficient wireless energy transfer in a multi-user MISO WIPT system, and investigated Quality of Service (QoS)-constrained robust beamforming to achieve the minimum transmission power. In addition, Ng. \emph{et. al.} \cite{DWKNg201309} also proposed a multi-objective approach to joint maximizing the energy harvesting efficiency and minimizing the total transmission power while ensuring secure communication in cognitive radio networks with WIPT.

\begin{figure}[htbp]
\centering
\includegraphics[width=3.in]{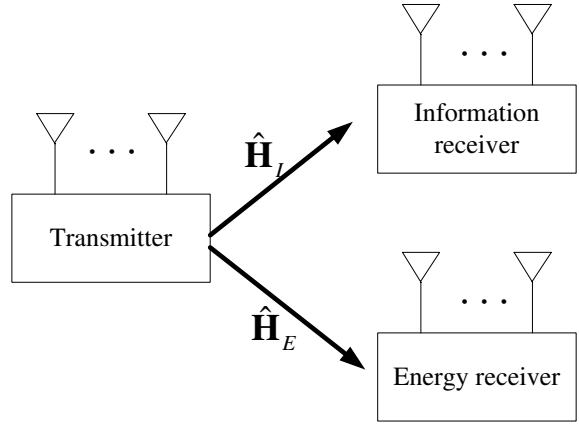}
\caption{The system model of a basic MIMO I-E broadcast system.}
\label{fig:fig0}
\end{figure}

Note that, none of existing works have investigated secure communications in MIMO WIPT systems. This paper considers a basic MIMO information-energy broadcast system as shown in Fig. \ref{fig:fig0}, where a multi-antenna transmitter transmits information and energy simultaneously to a multi-antenna information receiver and a multi-antenna energy receiver (or called energy harvester). We assume that the energy receiver is a dual-functional receiver which can also decode information from the received signal by switching its working mode from the EH mode to ID mode. Thus the energy receiver may eavesdrop the information intended for the information receiver only. By considering physical-layer security techniques, we study beamforming design to maximize the achievable secrecy rate subject to a total transmission power constraint and an energy harvesting constraint. The resulting secrecy rate maximization (SRM) problem is hard to solve due to not only the generally non-concave secrecy rate function but also the nonconvex EH constraint. First, we deal with the SRM problem by considering two special cases---the single-stream case and a specific full-stream case where the difference of Gram matrices of the channel matrices is positive semidefinite. For the two special cases, we propose global optimal solutions to the SRM problem based on semidefinite relaxation\cite{Luo2010}. Then, we treat the SRM problem of general case with arbitrary number of streams. We reformulate the SRM problem as another equivalent problem and propose inexact block coordinate descent (IBCD) algorithm to tackle the resulting problem. The convergence of the IBCD algorithm is studied in details. Furthermore, we extend the IBCD algorithm to the joint beamforming and artificial noise (AN) design problem. Finally, we evaluate the effectiveness of the proposed beamforming algorithms by simulations.

The remainder of this paper is organized as follows. In the next
section, we describe the problem formulation.
Section III presents global solutions to the single-stream case and a specific full-stream case, while Section IV proposes the IBCD algorithm to tackle the general case with an extension to joint beamforming and artificial noise design. In Section V we provide some numerical examples. Section VI concludes the paper.

\emph{Notations}: Throughout this paper, we use upper-case bold type for matrices,
lower--case bold type for column vectors, and regular type for scalars. For a square matrix $\bA$, $\bA^H$ denotes its Hermitian transpose, $\lambda_{\max}(\bA)$ ($\lambda_{\min}(\bA)$) denotes its maximum (minimum) eigenvalue, $\bA\succeq0$ ($\bA\nsucceq0$) represents that $\bA$ is (isn't) positive semidefinite, and $\bA\succ0$ denotes that $\bA$ is positive definite. $\bI$ denotes the identity matrix whose dimension will be clear from the context. The notations $\trace(\cdot)$, $\rank(\cdot)$ and $\det(\cdot)$ represent trace, rank and determinant operator, respectively. The distribution of a circularly symmetric complex Gaussian (CSCG) random vector with mean $\bm{\mu}$ and covariance matrix $\bC$ is denoted by $\cgauss(\bm{\mu},\bC)$, and `$\sim$' stands for `distributed as'. $\Cdom^{m\times n}$ denotes the space of $m\times n$ complex matrices. $\Re e\{a\}$ denotes the real part of a complex number $a$.

\section{System Model And Problem Formulation}
Consider an I-E broadcast system (see Fig. \ref{fig:fig0}) where one transmitter sends signal over the same spectrum to one IR and one ER with simultaneous information and power transfer. We assume that the transmitter is equipped with $N_T\geq 1$ antennas while the IR and ER are equipped with $N_{I}\geq 1$ and $N_{E}\geq 1$ antennas, respectively. Assuming a narrow-band transmission over the I-E broadcast system, the equivalent baseband channels from the transmitter to both receivers are modeled by
\begin{align}
&\by_{I} = \hH_{I}\bx+\bn_I,\\
&\by_{E} = \hH_{E}\bx+\bn_E
\end{align}
where $\by_{I}$ and $\by_{E}$ denote the received signal at the IR and ER, respectively, $\hH_{I}\in\Cdom^{N_I\times N_T}$  and $\hH_{E}\in\Cdom^{N_E\times N_T}$ denote the channel matrices from the transmitter to the IR and ER, respectively, $\bx\triangleq\bV\bs$ denotes the transmitted signal, $\bV\in\Cdom^{N_T\times d}$ is the transmit beamforming matrix employed by the transmitter, $\bs\sim\cgauss(0, \bI)$ denotes the transmitted symbols (a stream of length $d$) intended for the IR, $\bn_{I}\sim\cgauss(0, \sigma_I^2)$ and $\bn_{E}\sim\cgauss(0, \sigma_E^2)$ denote the additive white Gaussian noise (AWGN).

Furthermore, we assume that the ER can work in dual functions of information decoding and energy harvesting (i.e., either in ID mode or EH mode). In this scenario, the ER may potentially eavesdrop the information of the IR by switching its working mode to ID. To guarantee \emph{secure} transmission from the transmitter to the IR (no matter which mode the ER works in), the attractive physical layer security technique is assumed to be employed by the transmitter. Therefore, the achievable secrecy rate is given by\cite{Oggier2008}
\begin{equation}
\begin{split}
&C(\bV) \triangleq \log\det(\bI+\frac{1}{\sigma_I^2}\hH_I\bV\bV^H\hH_I^H)-\log\det(\bI+\frac{1}{\sigma_E^2}\hH_E\bV\bV^H\hH_E^H).
\end{split}
\end{equation}
On the other hand, the ER captures energy from the received signal $\by_E$. By neglecting the noise power, the harvested power at the ER is given by
\begin{equation}
E(\bV)\triangleq \zeta\trace(\hH_E\bV\bV^H\hH_E^H)
\end{equation}
where $0<\zeta\leq 1$ denotes the energy conversion efficiency.

In this paper, we are interested in beamforming design with the goal of maximizing the secrecy rate subject to both the harvested power constraint $E(\bV)\geq P_E$ and the total transmission power constraint $\trace(\bV\bV^H)\leq P_T$, where $P_T$ is the power budget for the transmitter and $P_E$ is the EH target required by the ER. For notational simplicity, we define $\bH_I\triangleq\frac{1}{\sigma_I}\hH_I$, $\bH_E\triangleq\frac{1}{\sigma_E}\hH_E$. The secrecy rate maximization problem can be stated as follows:
\begin{equation}\label{eq:P1_EH}
\begin{split}
 &\max_{\bV\in \Cdom^{N_t\times d}}~~\log\det(\bI+\bH_I\bV\bV^H\bH_I^H)-\log\det(\bI+\bH_E\bV\bV^H\bH_E^H)\\
&\st~ \trace(\bV\bV^H)\leq P_T,\\
&~~~~~\zeta\sigma_E^2\trace(\bH_E\bV\bV^H\bH_E^H)\geq P_{E}.
\end{split}
\end{equation}
Problem \eqref{eq:P1_EH} is feasible if and only if $\zeta \sigma_E^2P_T\lambda_{max}(\bH_E^H\bH_E)\geq P_E$. Furthermore, since the objective function of problem \eqref{eq:P1_EH} is generally not concave and the EH constraint is not convex, problem \eqref{eq:P1_EH} is nonconvex and hard to solve.
If we remove the EH constraint in problem \eqref{eq:P1_EH}, the resulting problem, denoted by $\mP_{noEH}$, is the beamforming design formulation of the well-known power-constrained SRM problem for Gaussian MIMO wiretap channel\cite{Shi2011,Li2011TIT,Liqiang2013}. It is known that\cite{Li2011TIT} problem $\mP_{noEH}$ must have positive maximum secrecy rate when $\bH_E^H\bH_E-\bH_I^H\bH_I\nsucceq 0$. However, this is not the case for problem \eqref{eq:P1_EH}. For example, consider the single-stream case with $d=1$. When $\zeta \sigma_E^2P_T\lambda_{max}(\bH_E^H\bH_E)=P_E$, problem \eqref{eq:P1_EH} has a unique feasible solution (up to phase rotation), for which, there exists some $\bH_I$ that achieves negative secrecy rate (given $\bH_E$) while satsifying $\bH_E^H\bH_E-\bH_I^H\bH_I\nsucceq 0$. Hence, we may obtain a negative maximum secrecy rate under the EH constraint even if $\bH_E^H\bH_E-\bH_I^H\bH_I\nsucceq 0$, which is not physically interesting. In this paper, \emph{we assume\footnote{If problem \eqref{eq:P1_EH} has a negative maximum secrecy rate, we may consider only secure information transmission by neglecting the EH constraint in practical implementation of the studied I-E broadcast system.} that problem \eqref{eq:P1_EH} has a positive optimal value}, and focus our efforts on algorithm design to tackle problem \eqref{eq:P1_EH}. It is worth mentioning that, to the best of our knowledge, problem  $\mP_{noEH}$ with $1<d<N_t$ has not yet considered in the literature. Moreover, the existing algorithms\cite{Shi2011,Li2011TIT,Liqiang2013} developed for problem $\mP_{noEH}$ with $d=1$ or $d=N_t$ don't apply to problem \eqref{eq:P1_EH}  due to the nonconvex EH constraint.

\section{Secure Beamforming Design: Global Solution To Two Special Cases}
In this section, we investigate problem \eqref{eq:P1_EH} by considering two special cases: single-stream case and full-stream case with $\bH_I^H\bH_I\succeq \bH_E^H\bH_E$. We propose global solutions to these two special cases.
\subsection{Single-stream case: $d=1$}
For the single-stream case, we below propose an optimal solution to problem \eqref{eq:P1_EH}.

In the single-stream case, the beamforming matrix $\bV$ reduces to a vector. For notational simplicity and clearance, we use $\bv$ to denote $\bV$ when $d=1$.
Using the identity\cite{Mtx_book} $\det\left(\bI+\bH\bv\bv^H\bH^H\right)=1+\bv^H\bH^H\bH\bv$, we can transform problem \eqref{eq:P1_EH} equivalently to
\begin{equation}\label{eq:P1_EH_eq}
\begin{split}
 &\max_{\bv}~~\frac{1+\bv^H\bH_I^H\bH_I\bv}{1+\bv^H\bH_E^H\bH_E\bv}\\
&\st~\bv^H\bv\leq P_T,\\
&~~~~~\bv^H\bH_E^H\bH_E\bv\geq \frac{P_{E}}{\zeta\sigma_E^2}.
\end{split}
\end{equation}
Problem \eqref{eq:P1_EH_eq} is a quadratically constrained quadratic fractional programming. By directly applying Charnes-Cooper transformation\cite{CCT1962,Chang2010} and semidefinite relaxation\cite{Luo2010} to problem \eqref{eq:P1_EH_eq}, we can turn the problem into a semidefinite programming (SDP) with three linear constraints and one additional variable (except the matrix variable). We below propose a more efficient solution to problem \eqref{eq:P1_EH_eq} by transforming the problem into a SDP with only two linear constraints.

Define $\bQ_E\triangleq\frac{1}{P_T}\bI+\bH_E^H\bH_E$, $\bQ_I\triangleq\frac{1}{P_T}\bI+\bH_I^H\bH_I$, $\bG\triangleq\frac{P_{E}}{P_T\zeta\sigma_E^2}\bI-\bH_E^H\bH_E$. We have Lemma \ref{eq:lemmad1}.
\begin{lemma}\label{eq:lemmad1}
\it{
Let $\bu^*$ be an optimal solution to the following problem
\begin{equation}\label{eq:P1_EH_eq_CC}
\begin{split}
 &\min_{\bu}~~\bu^H\bQ_E\bu\\
&\st~\bu^H\bQ_I\bu=1,\\
&~~~~~\bu^H\bG\bu\leq 0.
\end{split}
\end{equation}
Then $\bv^*=\sqrt{P_T}\frac{\bu^*}{\Vert\bu^*\Vert}$ is an optimal solution to problem \eqref{eq:P1_EH_eq}.}
\end{lemma}
The proof is relegated to Appendix A. Lemma \ref{eq:lemmad1} shows that the involving fractional form of the objective function of problem \eqref{eq:P1_EH_eq} can be removed without introducing extra quadratic constraints. Moreover, the optimal solution to problem \eqref{eq:P1_EH_eq} can be easily obtained as long as problem \eqref{eq:P1_EH_eq_CC} is solved.

Now let us consider how to solve problem \eqref{eq:P1_EH_eq_CC}.  We note that the second quadratic constraint of problem \eqref{eq:P1_EH_eq_CC} must be satisfied for any $\bu$ when $\zeta\sigma_E^2P_T\lambda_{min}(\bH_E^H\bH_E)\geq P_E$. In this case, problem \eqref{eq:P1_EH_eq_CC} simplifies to
\begin{equation}\label{eq:P1_EH_eq_simp}
\begin{split}
&\min \bu^H\bQ_E\bu\\
&\st~\bu^H\bQ_I\bu=1
\end{split}
\end{equation}
It is readily known\cite{Li2011TIT} that the eigenvector (with proper scaling) of the matrix $\bQ_E^{-1}\bQ_I$ corresponding to the minimum eigenvalue is an optimal solution to problem \eqref{eq:P1_EH_eq_simp}.

When $\zeta\sigma_E^2P_T\lambda_{min}(\bH_E^H\bH_E)< P_E$, problem \eqref{eq:P1_EH_eq_CC} is a quadratic optimization problem with two quadratic constraints. It is well-known\cite{Huang2010} that such a quadratic optimization problem can be globally solved using semidefinite relaxation. Defining $\bX=\bu\bu^H$ and ignoring the rank-one constraint, we obtain the SDR of problem \eqref{eq:P1_EH_eq_CC} as follows
\begin{equation}\label{eq:P1_EH_eq_SDR}
\begin{split}
 &\min_{\bX}~~\trace(\bQ_E\bX)\\
&\st~\trace(\bQ_I\bX)=1,\\
&~~~~~\trace(\bG\bX))\leq 0,\\
&~~~~~\bX\succeq 0.
\end{split}
\end{equation}
Problem \eqref{eq:P1_EH_eq_SDR} is a SDP which can be efficiently solved using interior-point algorithm\cite{cvx_book}. If the optimal solution to problem \eqref{eq:P1_EH_eq_SDR}, denoted by $\bX^*$, satisfies $\rank(\bX^*)=1$, then the optimal solution to problem \eqref{eq:P1_EH_eq_CC} can be obtained from the eigen-decomposition of $\bX^*$; otherwise, if $\rank(\bX^*)>1$, we run \emph{rank reduction procedure}\cite{Huang2010} to $\bX^*$ to get a rank-one solution to problem \eqref{eq:P1_EH_eq_SDR}  and then
perform eigen-decomposition on the rank-one solution to obtain the optimal solution to problem \eqref{eq:P1_EH_eq_CC}.

\subsection{Full-stream case: $d=N_T$ and $\bH_I^H\bH_I\succeq \bH_E^H\bH_E$}
When $d=N_T$, $\bV$ is a square matrix. In this case, by using SDR, i.e., defining $\bX=\bV\bV^H$ and dropping the rank constraint, we obtain the SDR of problem \eqref{eq:P1_EH}
\begin{equation}\label{eq:P1_EH_full}
\begin{split}
 &\max_{\bX}~~\log\det(\bI+\bH_I\bX\bH_I^H)-\log\det(\bI+\bH_E\bX\bH_E^H)\\
&\st~ \trace(\bX)\leq P_T,\\
&~~~~~\zeta\sigma_E^2\trace(\bH_E\bX\bH_E^H)\geq P_{E},\\
&~~~~~\bX\succeq 0.
\end{split}
\end{equation}
It is easily seen that the SDR is tight when $\bV$ is a square matrix and thus problem \eqref{eq:P1_EH_full} is equivalent to problem \eqref{eq:P1_EH}.
Moreover, it can be shown that the objective function of problem \eqref{eq:P1_EH_full} is concave when $\bH_I^H\bH_I\succeq \bH_E^H\bH_E$\cite{Shi2011,Oggier2008}. To facilitate using some off-the-shelf convex optimization tools (e.g., CVX\cite{cvx}), we below reformulate problem \eqref{eq:P1_EH_full} as an explicit convex problem with linear matrix inequality when $\bH_I^H\bH_I\succeq \bH_E^H\bH_E$.

\begin{prop}\label{prop:upperbound}
\it{
Problem \eqref{eq:P1_EH_full} is equivalent to the following convex problem
\begin{equation}\label{eq:SCM-eq}
\begin{split}
&\max_{\bX,\bY} ~\log\det\left(\bI+\bF^{\frac{1}{2}}\bY\bF^{\frac{1}{2}}\right)\\ 
&\st~\left[\begin{array}{cc}\bX-\bY&\bX\bH_E^H\\
\bH_E\bX&\bI+\bH_E\bX\bH_E^H
\end{array}\right]\succeq0,\\
&~~~~~ \trace(\bX)\leq P_T,\\
&~~~~~\zeta\sigma_E^2\trace(\bH_E\bX\bH_E^H)\geq P_{E},\\
&~~~~~\bX\succeq 0
\end{split}
\end{equation}}
where $\bF\triangleq\bH_I^H\bH_I-\bH_E^H\bH_E\succeq 0$.
\end{prop}
The proof is relegated to Appendix B. Since problem \eqref{eq:SCM-eq} is an explicit convex problem, it can be solved using, e.g., CVX.  Let $\tilde{\bX}$ be the optimal solution to problem \eqref{eq:SCM-eq}. The optimal solution to problem \eqref{eq:P1_EH} can be further obtained through eigen-decomposition of $\tilde{\bX}$.

\section{Secure Beamforming Design: KKT Solution To General Case}
In this section, we consider problem \eqref{eq:P1_EH} in the general case with arbitrary number of streams $d$. We first propose inexact block coordinate descent (IBCD) algorithm to tackle problem \eqref{eq:P1_EH} and then extend the IBCD algorithm to a more general case where artificial noise is employed to jam the energy harvester.

\subsection{Inexact Block Coordinate Algorithm For Problem \eqref{eq:P1_EH}}
Problem \eqref{eq:P1_EH} is generally much harder to solve than problem \eqref{eq:P1_EH_full} due to both the highly non-concave objective function and the nonconvex constraint $\zeta\sigma_E^2\trace(\bV^H\bH_E^H\bH_E\bV)\geq P_{E}$. To deal with the difficulties arising from the objective function and the nonconvex constraint, we first derive an equivalent problem of problem \eqref{eq:P1_EH} and then propose inexact block coordinate descent algorithm for the resulting problem.

\subsubsection{Reformulation of problem \eqref{eq:P1_EH}}
To tackle the difficulty arising from the Shannon capacity expression in the objective function of problem \eqref{eq:P1_EH}, we extend the key idea of the popular WMMSE algorithm\cite{Shi2011,Christensen2008}, which is commonly used to address rate/sum-rate maximization problems, to reformulating problem \eqref{eq:P1_EH}. The key idea behind the WMMSE algorithm is transforming a rate or sum-rate maximization problem to another equivalent problem (by introducing auxiliary variables) which allows using simple block coordinate decent method\cite{Bertsekas_book}. Such an idea is based on three important facts which are summarized in Lemma \ref{lem:key_idea_WMMSE}.
\begin{lemma}\label{lem:key_idea_WMMSE}
\it{
Define an $m$ by $m$ matrix function $$\bE(\bU, \bV) \triangleq(\bI-\bU^H\bH\bV)(\bI-\bU^H\bH\bV)^H+\bU^H\bN\bU$$ where $\bN$ is any positive definite matrix. The following three facts hold true.

\begin{itemize}
\item [1)] For any positive definite matrix $\bE\in\Cdom^{m\times m}$, we have
\begin{equation}
\bE^{-1}=\arg\max_{\bW\succ0}\log\det(\bW)-\trace(\bW\bE)
\end{equation}
and
\begin{equation}
-\log\det(\bE)=\max_{\bW\succ0}\log\det(\bW)-\trace(\bW\bE)+m.
\end{equation}
\item [2)]  For any positive definite matrix $\bW$, we have
\begin{equation}
\begin{split}
\tilde{\bU}&\triangleq\arg\min_{\bU}\trace(\bW\bE(\bU, \bV))\\
&=\left(\bN+\bH\bV\bV^H\bH^H\right)^{-1}\bH\bV
\end{split}
\end{equation}
and
\begin{equation}\label{eq:MSE}
\begin{split}
\bE(\tilde{\bU}, \bV)
=&\bI-\tilde{\bU}^H\bH\bV\\
=&\left(\bI+\bV^H\bH^H\bN^{-1}\bH\bV\right)^{-1}.
\end{split}
\end{equation}
\item [3)] We have
\begin{equation}
\begin{split}
&\log\det(\bI+\bH\bV\bV^H\bH^H\bN^{-1})\\
=&\max_{\bW\succ0, \bU}\log\det(\bW)-\trace(\bW\bE(\bU,\bV))+m.
\end{split}
\end{equation}
\end{itemize}}
\end{lemma}
Facts 1) and 2) can be proven by simply using the first-order optimality condition, while Fact 3) directly follows from Facts 1) and 2) and the identity $\log\det(\bI+\bA\bB)=\log\det(\bI+\bB\bA)$. We refer readers to \cite{Shi2011,Christensen2008} for more detailed proof.

Next, using Lemma \ref{lem:key_idea_WMMSE}, we derive an equivalent problem of problem \eqref{eq:P1_EH} by introducing some auxiliary variables. Define
\begin{equation}
\mE(\bU, \bV)\triangleq(\bI-\bU^H\bH_I\bV)(\bI-\bU^H\bH_I\bV)^H+\bU^H\bU.
\end{equation}
Then we have from Fact 3) that
\begin{equation}\label{eq:obj1}
\begin{split}
&\log\det(\bI+\bH_I\bV\bV^H\bH_I^H)\\
=&\max_{\bW_I\succ0, \bU}\log\det(\bW_I)-\trace(\bW_I\mE(\bU,\bV))+d
\end{split}
\end{equation}
Furthermore, from Fact 1), we have
\begin{equation}\label{eq:obj2}
\begin{split}
&-\log\det(\bI+\bH_E\bV\bV^H\bH_E^H)\\
=&\max_{\bW_E\succ0}\log\det(\bW_E)-\trace\left(\bW_E\left(\bI+\bH_E\bV\bV^H\bH_E^H\right)\right)+N_E
\end{split}
\end{equation}
Since the objective function of problem \eqref{eq:P1_EH} is equivalent to the sum of the right-hand-side (RHS) of \eqref{eq:obj1} and \eqref{eq:obj2}, problem \eqref{eq:P1_EH} is equivalent to
\begin{equation}\label{eq:P1_EH_WMMSE}
\begin{split}
 &\max_{\bV,\bW_I\succ0,\bW_E\succ0, \bU}~~\log\det(\bW_I)-\trace(\bW_I\mE(\bU, \bV))+d\\
&~~+\log\det(\bW_E)-\trace\left(\bW_E\left(\bI+\bH_E\bV\bV^H\bH_E^H\right)\right)+N_E\\
&\st~ \trace(\bV\bV^H)\leq P_T,\\
&~~~~~\zeta\sigma_E^2\trace(\bV^H\bH_E^H\bH_E\bV)\geq P_{E}.
\end{split}
\end{equation}

\subsubsection{Inexact block coordinate descent algorithm for problem \eqref{eq:P1_EH_WMMSE}}
Although problem \eqref{eq:P1_EH_WMMSE} has more variables than problem \eqref{eq:P1_EH}, the former allows using simple block coordinate decent method, which optimizes the objective function over one variable (or one group of variables) while keeping all the others fixed at a time. In the BCD method applied to problem \eqref{eq:P1_EH_WMMSE}, it is required to iteratively solve three (or four) subproblems, among which, the most difficult one is
\begin{equation}\label{eq:subpV}
\begin{split}
&\min_{\bV} \trace(\bV^H\bH_I^H\bU\bW_I\bU^H\bH_I\bV)-\trace(\bW_I\bU^H\bH_I\bV)\\
&~~~~~~~~-\trace(\bW_I\bV^H\bH_I^H\bU)+\trace(\bV^H\bH_E^H\bW_E\bH_E\bV)\\
&\st~ \trace(\bV\bV^H)\leq P_T,\\
&~~~~~\zeta\sigma_E^2\trace(\bV^H\bH_E^H\bH_E\bV)\geq P_{E}
\end{split}
\end{equation}
which is obtained from \eqref{eq:P1_EH_WMMSE} by fixing $\bW_I$, $\bW_E$ and $\bU$.
Note that, problem \eqref{eq:subpV} is a nonconvex problem due to the nonconvex EH constraint. To globally solve it, we need to use SDR and rank-one reduction technique\footnote{Denote by $\bv$ the vectorization of the matrix variable $\bV$. We first reformulate problem \eqref{eq:subpV} as a quadratic optimization problem with respect to variable $\bv$. Then, by defining $\bZ\triangleq[\bv^H~ 1]^H [\bv^H~ 1]$ and relaxing the rank-one constraint, we can relax the resulting quadratic optimization problem as an SDP. Finally, from the SDR solution, we can find the optimal solution to problem \eqref{eq:subpV} by performing eigen-decomposition and rank-one reduction\cite{Huang2010}.}. As a result, the BCD method applied to problem \eqref{eq:P1_EH_WMMSE} requires solving a number of semidefinite programmings as the iterations proceed, which makes the algorithm less efficient. For better efficiency and also ease of implementation, we propose \emph{inexact block coordinate descent} (IBCD) method to tackle problem \eqref{eq:P1_EH_WMMSE}.

Similar to the BCD method, the IBCD method iteratively updates one (or one group of) variable while fixing the others.
However, in the IBCD method, it is not required to globally solve all the subproblems; instead, we only find an \emph{inexact} solution to some subproblems while keeping the objective function non-descending. Specifically, each iteration of the IBCD method consists of the following three sub-iterations.

\textbf{\emph{Sub-iteration 1:}} Solve \eqref{eq:P1_EH_WMMSE} for $\bU$ while fixing $\bV$, $\bW_I$ and $\bW_E$. This is equivalent to minimizing $\trace(\bW_I\mE(\bU, \bV))$ over $\bU$. According to Fact 2) in Lemma \ref{lem:key_idea_WMMSE}, we obtain the optimal $\bU$ given $\bV$ as follows
\begin{equation}
\bU = \left(\bI+\bH_I\bV\bV^H\bH_I^H\right)^{-1}\bH_I\bV.
\end{equation}

\textbf{\emph{Sub-iteration 2:}}  Solve \eqref{eq:P1_EH_WMMSE} for $\bW_I$ and $\bW_E$ while fixing $\bU$ and $\bV$. Note that the objective function of problem \eqref{eq:P1_EH_WMMSE} is separable over $\bW_I$ and $\bW_E$. Hence, Using Fact 1) in Lemma \ref{lem:key_idea_WMMSE} twice, we can easily obtain the optimal $\bW_I$ and $\bW_E$ given $\bU$ and $\bV$ as follows
\begin{align}
&\bW_I = \mE(\bU, \bV)^{-1}\label{eq:update_W_I}\\
&\bW_E = \left(\bI+\bH_E\bV\bV^H\bH_E^H\right)^{-1}
\end{align}

\textbf{\emph{Sub-iteration 3:}} To update $\bV$ while fixing $\bW_I$, $\bW_E$, $\bU$, we solve the following subproblem (instead of problem \eqref{eq:subpV} in the BCD method):
\begin{equation}\label{eq:subpV-linearization}
\begin{split}
&\min_{\bV} \trace(\bV^H\bH_I^H\bU\bW_I\bU^H\bH_I\bV)-\trace(\bW_I\bU^H\bH_I\bV)\\
&~~~~~~~~-\trace(\bW_I\bV^H\bH_I^H\bU)+\trace(\bV^H\bH_E^H\bW_E\bH_E\bV)\\
&\st~ \trace(\bV\bV^H)\leq P_T,\\
&~~~~~\trace(\tilde{\bV}^H\bH_E^H\bH_E\tilde{\bV})
+\trace(\tilde{\bV}^H\bH_E^H\bH_E(\bV-\tilde{\bV}))\\
&~~~~~~~~~~+\trace((\bV-\tilde{\bV})^H\bH_E^H\bH_E\tilde{\bV})\geq \frac{P_{E}}{\zeta\sigma_E^2}
\end{split}
\end{equation}
Problem \eqref{eq:subpV-linearization} is obtained by replacing the quadratic function $\trace(\bV^H\bH_E^H\bH_E\bV)$ in the EH constraint of problem \eqref{eq:subpV} with its first-order approximation at $\tilde{\bV}$, where $\tilde{\bV}$ is the update of $\bV$ obtained in the last iteration. In contrast to problem \eqref{eq:subpV}, problem \eqref{eq:subpV-linearization} admits an efficient solution. As it will be shown later, although the solution to problem \eqref{eq:subpV-linearization} is just a feasible solution to problem \eqref{eq:subpV}, it can keep the objective function of problem \eqref{eq:P1_EH_WMMSE} non-descending.

\underline{\emph{Solution to problem \eqref{eq:subpV-linearization}}}: we here show how problem \eqref{eq:subpV-linearization} can be solved efficiently. Note that problem \eqref{eq:subpV-linearization} is a convex problem. Thus, it can be solved by dealing with its dual problem. To this end, by introducing Lagrange multiplier $\lambda$ for the first constraint of problem
\eqref{eq:subpV-linearization}, we define the partial Lagrangian associated with problem \eqref{eq:subpV-linearization} as
\begin{equation}
\begin{split}
&\mathcal{L}(\bV, \lambda)\triangleq \trace(\bV^H\bH_I^H\bU\bW_I\bU^H\bH_I\bV)-\trace(\bW_I\bU^H\bH_I\bV)\\
&~~~~~~~~-\trace(\bW_I\bV^H\bH_I^H\bU)+\trace(\bV^H\bH_E^H\bW_E\bH_E\bV)\\
&~~~~~~~~+\lambda\left(\trace(\bV\bV^H)-P_T\right).
\end{split}
\end{equation}
Furthermore, we define $b(\tV)\triangleq\frac{P_{E}}{\zeta\sigma_E^2}+\trace(\tilde{\bV}^H\bH_E^H\bH_E\tilde{\bV})$. Thus the second constraint of problem \eqref{eq:subpV-linearization} can be compactly written as
$$2\Re e\left\{\trace\left(\bV^H\bH_E^H\bH_E\tV\right)\right\}\geq b(\tV).$$
and the dual problem of problem \eqref{eq:subpV} is
\begin{equation}
\begin{split}
&\max_{\lambda}h(\lambda)\\
&\st~\lambda\geq 0
\end{split}
\end{equation}
where $h(\lambda)$ is the dual function given by
\begin{equation}\label{eq:dualfun}
\begin{split}
h(\lambda)\triangleq&\min_{\bV}\mathcal{L}(\bV, \lambda)\\
&\st~2\Re e\left\{\trace\left(\bV^H\bH_E^H\bH_E\tV\right)\right\}\geq b(\tV).
\end{split}
\end{equation}
Note that problem \eqref{eq:dualfun} is a linearly constrained convex quadratic optimization problem. It can be solved in closed-form by using Lagrange multiplier method. The solution to problem \eqref{eq:dualfun} given\footnote{If the solution to problem \eqref{eq:dualfun} with $\lambda=0$ satisfies the total power constraint, then the optimal $\lambda$ is zero.} $\lambda>0$ is summarized in Proposition \ref{prop:dualfun_sol}.
\begin{prop}\label{prop:dualfun_sol}
\it{
Let $\bP\bSigma\bP^H$ be the eigen-decomposition of the matrix $\bH_I^H\bU\bW_I\bU^H\bH_I+\bH_E^H\bW_E\bH_E$ where $\bP$ consists of the orthonormal eigenvectors and $\bSigma$ is a diagonal matrix with each diagonal entry being the corresponding eigenvalue.
Define
$\bTheta(\lambda)\triangleq\bP\left(\lambda\bI+\bSigma\right)^{-1}\bP^H$. Given $\lambda>0$, the optimal solution to problem \eqref{eq:dualfun} can be expressed as
\begin{equation}\label{eq:dualfun_sol}
\bV^*=\bTheta(\lambda)\left(\bH_I^H\bU\bW_I+\mu^*\bH_E^H\bH_E\tilde{\bV}\right)
\end{equation}
where
\begin{equation}
\mu^*=\frac{\max\left(b(\tV)-2\Re e\left\{\trace\left(\tilde{\bV}^H\bH_E^H\bH_E\bTheta(\lambda)\bH_I^H\bU\bW_I\right)\right\}, 0\right)}{2\trace\left(\tilde{\bV}^H\bH_E^H\bH_E\bTheta(\lambda)\bH_E^H\bH_E\tilde{\bV}\right)}
\end{equation}
Moreover, $\trace\left(\bV^*(\bV^*)^H\right)-P_T$ is the derivative of $h(\lambda)$.}
\end{prop}
The proof is easy and the details are omitted for brevity. Eq. \eqref{eq:dualfun_sol} is obtained by using Lagrange multiplier method with $\mu^*$ being the optimal Lagrange multiplier associated with the linear constraint. Note that $\mu^*=0$ corresponds to the case when the solution to the unconstrained version of problem \eqref{eq:dualfun} satisfies the linear constraint. Furthermore, since the objective function of problem \eqref{eq:dualfun} given $\lambda>0$ is strictly convex, problem \eqref{eq:dualfun} has a unique solution. It follows that $h(\lambda)$ is differentiable for $\lambda>0$ and its derivative is simply $\trace\left(\bV^*(\bV^*)^H\right)-P_T$. With this analytic form derivative, the dual problem (equivalently, problem \eqref{eq:subpV-linearization}) can be efficiently solved using Bisection method\cite{cvx_book}, which is summarized in TABLE I.
\begin{table}[htbp]
\centering
\caption{Pseudo code of Bisection method for problem \eqref{eq:subpV-linearization}} \label{fig:pseudo_bisec}
\begin{tabular}{|p{3in}|}
\hline
\begin{itemize}
\item [1] \; Initialize $0\leq\lambda_l<\lambda_u$
\item [2] \; \textbf{repeat}
\item [3]\; \quad $\lambda\leftarrow \frac{\lambda_l+\lambda_u}{2}$
\item [4] \quad $\mu \leftarrow \frac{\max\left(b(\tV)-2\Re e\left\{\trace\left(\tilde{\bV}^H\bH_E^H\bH_E\bTheta(\lambda)\bH_I^H\bU\bW_I\right)\right\}, 0\right)}{2\trace\left(\tilde{\bV}^H\bH_E^H\bH_E\bTheta(\lambda)\bH_E^H\bH_E\tilde{\bV}\right)}$
\item [5] \quad $\bV \leftarrow \bTheta(\lambda)\left(\bH_I^H\bU+\mu\bH_E^H\bH_E\tilde{\bV}\right)$
\item [6] \quad \textbf{If} $\trace(\bV\bV^H)-P_T\geq 0$
\item [7] \quad\quad $\lambda_l \leftarrow \lambda$
\item [8] \quad\textbf{else}
\item [9] \quad\quad $\lambda_u \leftarrow \lambda$
\item [10] \quad \textbf{end}
\item [11] \; \textbf{until} $\left|\lambda_u-\lambda_l\right|\leq \epsilon$
\end{itemize}
\\
\hline
\end{tabular}
\end{table}

\begin{table}[htbp]
\centering
\caption{Pseudo code of the IBCD method for problem \eqref{eq:P1_EH}} \label{fig:pseudo_code_MIMO}
\begin{tabular}{|p{3.in}|}
\hline
\begin{itemize}
\item [1] \; Initialize $\bV$'s such that $\trace\left(\bV\bV^H\right) = P_T$ and $\zeta\sigma_E^2\trace(\bV^H\bH_E^H\bH_E\bV)\geq P_{E}$
\item [2] \; \textbf{repeat}
\item [3]\; \quad $\tV\leftarrow \bV$
\item [4] \quad $\bU \leftarrow \left(\bI+\bH_I\tV\tV^H\bH_I^H\right)^{-1}\bH_I\tV$
\item [5] \quad  $\bW_I \leftarrow \bI+\tV^H\bH_I^H\bH_I\tV$
\item [6] \quad $\bW_E \leftarrow \left(\bI+\bH_E\tV\tV^H\bH_E^H\right)^{-1}$
\item [7] \quad update $\bV$ by solving problem \eqref{eq:subpV-linearization} using Bisection method.
\item [8] \; \textbf{until} $\left|C(\bV)-C(\tV)\right|\leq \epsilon$
\end{itemize}
\\
\hline
\end{tabular}
\end{table}
Finally, we summarize the proposed algorithm\footnote{It is readily known that the complexity of the proposed algorithm is dominated by the eigen-decomposition operation. Assuming $N_T\geq \max(N_I, N_E)$, it can be shown that each iteration of the proposed IBCD method has complexity of $O(N_T^3)$. However, if we use SDR to directly solve problem \eqref{eq:subpV}, the complexity is at least $O(d^{3.5}N_T^{3.5})$\cite{cvx_book}.} for problem
\eqref{eq:P1_EH} in TABLE I, where Steps 4-7 correspond to the three sub-iterations of the IBCD method. Note that Step 5 follows from \eqref{eq:update_W_I} and \eqref{eq:MSE}. The following proposition summarizes the convergence property of the IBCD method.
\begin{prop}\label{prop:stationary}
\it{
The IBCD algorithm produces non-descending objective value sequence. Moreover, every limit point $(\bU^*, \bV^*, \bW_I^*, \bW_E^*)$ of the iterates generated by the IBCD algorithm is a KKT point of problem \eqref{eq:P1_EH_WMMSE}, and the corresponding $\bV^*$ is a KKT point of problem \eqref{eq:P1_EH}.
}
\end{prop}
The proof of Proposition \eqref{prop:stationary} is relegated to Appendix C. It indicates that the proposed algorithm monotonically converges to a stationary point of problem \eqref{eq:P1_EH}. The monotonic convergence is attractive since it guarantees an improved objective value with arbitrary random initialization. The convergence performance of the IBCD method is further explored later with numerical examples.

\subsection{Extension To Joint Artificial Noise and Beamforming Design}
We here consider an extension of the IBCD algorithm to the case where the transmitter also sends artificial noise\footnote{The IBCD algorithm can be also extended to the AN plus energy beamforming case.} (AN) to jam the energy harvester in order to achieve better secrecy rate\cite{Liao2011,Liqiang2013}. In this case, the transmitted signal is expressed as $\bx\triangleq\bV\bs+\bn$ where $\bn$ represents the artificial noise with zero mean and covariance matrix $\bZ$. The achievable secrecy rate is given by
\begin{equation}\label{eq:SR-AN}
\begin{split}
&C_{AN}(\bV, \bZ) \triangleq \log\det\left(\bI+\bH_I\bV\bV^H\bH_I^H(\bI+\bH_I\bZ\bH_I^H)^{-1}\right)\\
&~~~~~~-\log\det\left(\bI+\bH_E\bV\bV^H\bH_E^H(\bI+\bH_E\bZ\bH_E^H)^{-1}\right).
\end{split}
\end{equation}
The corresponding secrecy rate maximization problem is stated as
\begin{equation}\label{eq:SRP-AN}
\begin{split}
&\max_{\bV, \bZ}~ C_{AN}(\bV, \bZ)\\
&\st~\trace(\bV\bV^H+\bZ)\leq P_T,\\
&~~~~~\trace\left(\bH_E(\bV\bV^H+\bZ)\bH_E^H\right)\geq \frac{P_E}{\zeta\sigma_E^2},\\
&~~~~~\bZ\succeq 0.
\end{split}
\end{equation}
By variable substitution $\bZ=\bV_E\bV_E^H$ with $\bV_E\in \Cdom^{N_t\times N_t}$, problem \eqref{eq:SRP-AN} is equivalent to
\begin{equation}\label{eq:SRP-AN-eq}
\begin{split}
&\max_{\bV, \bV_E} C_{AN}(\bV, \bV_E\bV_E^H)\\
&\st~\trace(\bV\bV^H+\bV_E\bV_E^H)\leq P_T,\\
&~~~~~\trace\left(\bH_E(\bV\bV^H+\bV_E\bV_E^H)\bH_E^H\right)\geq \frac{P_E}{\zeta\sigma_E^2}.
\end{split}
\end{equation}

Next, we derive an equivalent problem of problem \eqref{eq:SRP-AN-eq}. First, we have
\begin{equation}
\begin{split}
C_{AN}(\bV, &\bV_E\bV_E^H)=\underbrace{\log\det(\bI+\bH_I\bV\bV^H\bH_I^H(\bI+\bH_I\bV_E\bV_E^H\bH_I^H)^{-1})}_{f_1}\\
&+\underbrace{\log\det(\bI+\bH_E\bV_E\bV_E^H\bH_E^H)}_{f_2}-\underbrace{\log\det(\bI+\bH_E\bV_E\bV_E^H\bH_E^H+\bH_E\bV\bV^H\bH_E^H)}_{f_3}.
\end{split}
\end{equation}
Furthermore, according to Lemma \ref{lem:key_idea_WMMSE}, we have
\begin{align}
&f_1=\max_{\bW_1\succ0, \bU_1}\log\det(\bW_1)-\trace(\bW_1\mE_1(\bU_1, \bV, \bV_E))+d,\\
&f_2=\max_{\bW_2\succ0, \bU_2}\log\det(\bW_2)-\trace(\bW_2\mE_2(\bU_2, \bV_E))+N_t,\\
&f_3=\max_{\bW_3\succ0}\log\det(\bW_3)-\trace\left(\bW_3(\bI+\bH_E\bV_E\bV_E^H\bH_E^H+\bH_E\bV\bV^H\bH_E^H)\right)+N_E,
\end{align}
where
\begin{align}
&\mE_1(\bU_1, \bV, \bV_E)\triangleq(\bI-\bU_1^H\bH_I\bV)(\bI-\bU_1^H\bH_I\bV)^H+\bU_1^H(\bI+\bH_I\bV_E\bV_E^H\bH_I^H)\bU_1,\\
&\mE_2(\bU_2, \bV_E)\triangleq\left(\bI-\bU_2^H\bH_E\bV_E\right)\left(\bI-\bU_2^H\bH_E\bV_E\right)^H+\bU_2^H\bU_2.
\end{align}
Therefore, the secrecy rate maximization problem in the AN case is equivalent to
\begin{equation}\label{eq:SR-AN-WMMSE}
\begin{split}
&\max_{\bW_1\succ0,\bW_2\succ0,\bW_3\succ0, \bU_1, \bU_2, \bV, \bV_E}~\log\det(\bW_1)-\trace(\bW_1\mE_1(\bU_1, \bV, \bV_E))\\
&~~~~~~~+\log\det(\bW_2)-\trace(\bW_2\mE_2(\bU_2, \bV_E))\\
&~~~~~~~+\log\det(\bW_3)-\trace\left(\bW_3(\bI+\bH_E\bV_E\bV_E^H\bH_E^H+\bH_E\bV\bV^H\bH_E^H)\right),\\
&\st~\trace(\bV\bV^H+\bV_E\bV_E^H)\leq P_T,\\
&~~~~~~\trace\left(\bH_E(\bV\bV^H+\bV_E\bV_E^H)\bH_E^H\right)\geq \frac{P_E}{\zeta\sigma_E^2}.
\end{split}
\end{equation}


The IBCD algorithm can be generalized to tackle problem \eqref{eq:SR-AN-WMMSE}. In each iteration, given $\bV$ and $\bV_E$, we can update $\bU_1$, $\bU_2$, $\bW_1$, $\bW_2$, $\bW_3$ in closed-form,  respectively, while, to update $\bV$ and $\bV_E$ given $(\bU_1, \bU_2, \bW_1, \bW_2, \bW_3)$, we can linearize the EH constraint and solve the resulting problem using Bisection method. Furthermore,  we can similarly prove that the algorithm can monotonically converge to a KKT point of problem \eqref{eq:SRP-AN}.


\section{Numerical Examples}
In this section, we provide numerical examples to illustrate the performance of the proposed beamforming algorithms. In all our simulations, we assume that both the IR and ER are equipped with two antennas. Moreover, we set $\sigma_I^2=\sigma_E^2=-50$dBm and $\zeta=0.5$. It is further assumed that the signal attenuation from the transmitter to both receivers is $50$dB corresponding to an identical distance of about $5$ meters. The channels from the transmitter to both receivers are randomly generated from i.i.d Rayleigh fading with the average power specified as above (i.e., 1e-5). It should be noted that, since the IBCD algorithm requires feasible initialization, we run a warmstart procedure to obtain an efficient feasible initial point. The warmstart procedure consists of the following three steps: 1) randomly generate $\bV$; 2) update $\bU$, $\bW_I$ and $\bW_E$ as in Steps 4-6 in TABLE II; 3) obtain a feasible $\bV$ by solving \eqref{eq:subpV} using SDR as argued in footnote 3. A similar warmstart procedure is also performed for the beamforming algorithm with artificial noise.

\begin{figure}[htbp]
\centering
\includegraphics[width=3.in]{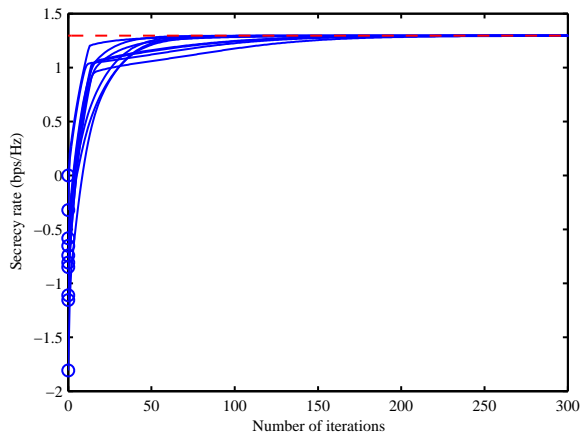}
\caption{The secrecy rate Vs. iterations: single-stream case.}
\label{fig:fig1}
\end{figure}

\subsection{Convergence performance}
\begin{figure}[htbp]
\centering
\includegraphics[width=3.in]{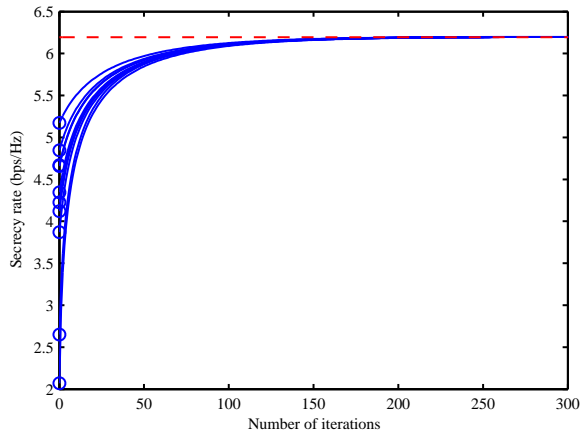}
\caption{The secrecy rate Vs. iterations: full-stream case.}
\label{fig:fig2}
\end{figure}

First, we investigate the convergence performance of the IBCD algorithm for problem \eqref{eq:P1_EH} by comparing with the global solutions in two special cases. We first consider the single-stream case with $P_T=10$dBm, $P_E=-40$dBm, and $N_T=4$. An example of convergence behavior of the IBCD algorithm is shown in Fig. \ref{fig:fig1}, where circles represent different initializations and the dotted horizontal line denotes the optimal value obtained by the SDR method in Section III.A. It is observed that the IBCD algorithm can converge to the global optimal solution irrespective of initial points. We then consider the full-stream case with $P_T=20$dBm, $P_E=-30$dBm, and
\begin{equation*}
\bH_I = \left[\begin{array}{cc}
   -0.8355{-}0.4547i & 1.5249{+}0.9305i\\
   1.1033{-}0.9940i &  1.6232 {-}1.0196i\\
\end{array}\right],
\end{equation*}
\begin{equation*}
\bH_E = \left[\begin{array}{cc}
   0.1409 {-} 0.1914i & 0.3241{+}0.2328i\\
   0.7981 {+} 0.7771i &-0.9295 {+} 0.0945i
\end{array}\right].
\end{equation*}
It can be easily verified that the matrix $\bH_I^H\bH_I-\bH_E^H\bH_E$ is positive semidefinite. Hence, the optimal value of problem \eqref{eq:P1_EH} in this case can be obtained by the proposed method in Section III.B. Figure \ref{fig:fig2} shows the corresponding convergence performance of the IBCD algorithm, where the dotted horizontal line represents the optimal value. As in the single-stream case, it is observed that the IBCD algorithm has global convergence.

\begin{figure}[htbp]
\centering
\includegraphics[width=3.in]{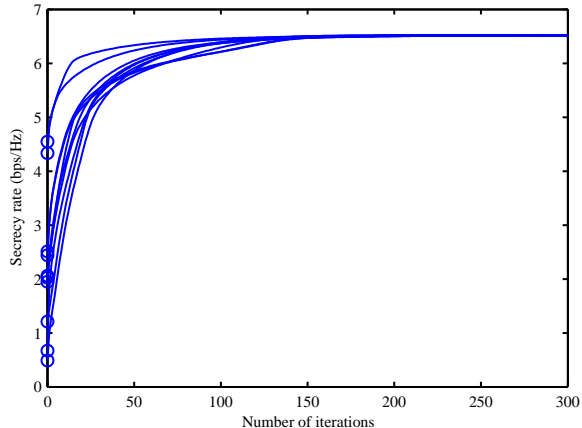}
\caption{The secrecy rate Vs. iterations: artificial noise case.}
\label{fig:fig3}
\end{figure}

Then, we demonstrate the convergence performance of the generalized IBCD algorithm for problem \eqref{eq:SRP-AN}. Figure \ref{fig:fig3} shows an example of convergence behavior of the generalized IBCD algorithm with $P_T=15$dBm, $P_E=-35$dBm, and $N_T=4$. It is seen that the generalized IBCD algorithm finally reaches the same objective value of problem  \eqref{eq:SRP-AN} regardless of initial points.

To summarize, the above numerical examples indicate that the IBCD algorithm has good convergence performance although both problem \eqref{eq:P1_EH} and problem \eqref{eq:SRP-AN} are highly nonconvex.

\subsection{Secrecy rate performance}
In this set of simulations, we investigate the secrecy rate performance of the proposed beamforming algorithms with/without artificial noise. We set the number of streams $d$ to be $2$. For both the (generalized) IBCD algorithm and the Bisection algorithm, we set $\epsilon=1e-6$ to achieve a good accuracy.
\begin{figure}[htbp]
\centering
\includegraphics[width=3.in]{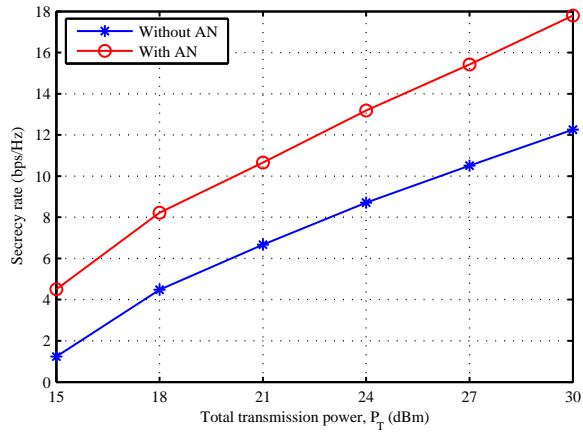}
\caption{The secrecy rate Vs. total transmission power.}
\label{fig:fig4}
\end{figure}

First, we investigate the achieved secrecy rate versus the total transmission power, with the harvested power target, $P_E$, being fixed as $-30$dBm. It is assumed that the transmitter is equipped with $N_T=4$ antennas. Figure \ref{fig:fig4} shows the achieved secrecy rate of the beamforming algorithms with and without artificial noise, where each data point is averaged over 100 random channel realizations. It is observed that the achieved secrecy rate increases with the total transmission power. Furthermore, it is seen that better secrecy rate can be achieved with the aid of artificial noise.

\begin{figure}[htbp]
\centering
\includegraphics[width=3.in]{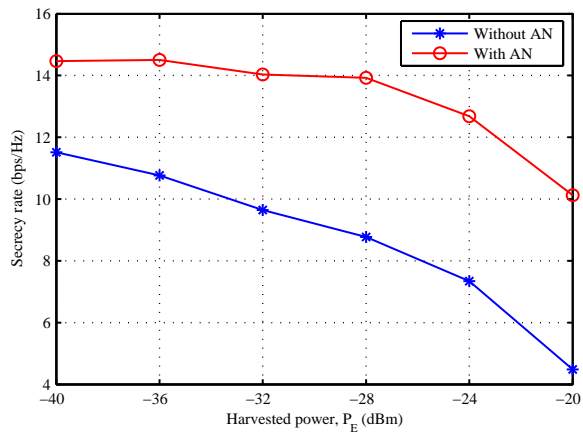}
\caption{The secrecy rate Vs. harvested power.}
\label{fig:fig5}
\end{figure}

Next, we show in Fig. \ref{fig:fig5} the achieved secrecy rate versus the harvested power $P_E$ with fixed $P_T=25$dBm and $N_T=4$. It is observed that, for both artificial noise case and no artificial noise case, the secrecy rate decreases as the harvested power target increases. Moreover, similarly as in Fig. \ref{fig:fig4}, it is seen that the AN-aided beamforming design method outperforms the beamforming design method without AN in terms of the achieved secrecy rate.

\begin{figure}[htbp]
\centering
\includegraphics[width=3.in]{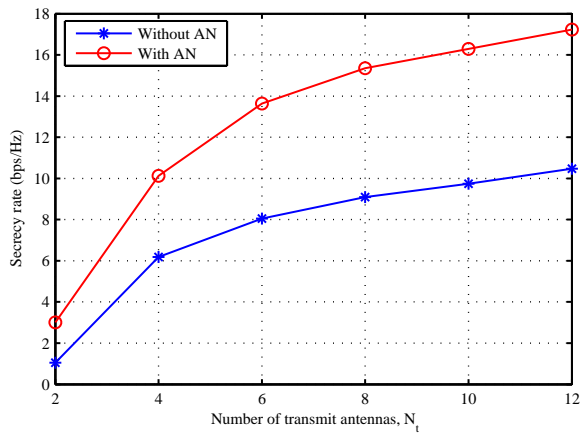}
\caption{The secrecy rate Vs. number of transmit antennas.}
\label{fig:fig6}
\end{figure}

At last, we plot the secrecy rate achieved by the proposed beamforming algorithms versus the number of transmit antennas in Fig. \ref{fig:fig6} with fixed $P_T=20$dBm and $P_E=-30$dBm. Again, it is observed that the AN-aided beamforming design method achieves better secrecy rate performance. However, the gap between the secrecy rate of the beamforming algorithms with and without artificial noise increases with the number of transmit antennas. This indicates that the artificial noise could impose more positive impact on the secrecy rate when the number of transmit antennas is large.

\section{Conclusions}
This paper has studied secure beamforming design for a two-user MIMO information-energy broadcast system. The problem of secrecy rate maximization subject to an energy harvesting constraint and a total power constraint is investigated. First, global optimal beamforming solutions are proposed for both the single-stream case and the full-stream case with channels satisfying positive semidefiniteness. Then, by developing the IBCD algorithm, a simple iterative beamforming solution is proposed for the general case with arbitrary number of streams. It is proven that the IBCD algorithm has monotonic convergence and any limit point of the IBCD algorithm is a KKT solution to the studied secrecy rate maximization problem. Furthermore, the IBCD algorithm is generalized to joint beamforming and artificial noise design. Finally, simulation results show that better secrecy rate is achieved with the aid of artificial noise.

\appendices
\section{The Proof of Lemma \ref{eq:lemmad1}}
By assumption (i.e., problem \eqref{eq:P1_EH} has positive optimal value), we have $1+\bv^H\bH_I^H\bH_I\bv\geq1+\bv^H\bH_E^H\bH_E\bv$ at the optimality of problem \eqref{eq:P1_EH_eq}. On the other hand, it is known that the objective function of problem \eqref{eq:P1_EH_eq} is an increasing function in $\vert\bv\vert$ if $1+\bv^H\bH_I^H\bH_I\bv\geq1+\bv^H\bH_E^H\bH_E\bv$. Hence, the total power constraint must be active at the optimality of problem \eqref{eq:P1_EH_eq}. It follows that problem \eqref{eq:P1_EH_eq} has the same optimal solution set as that of the following problem
\begin{equation}\label{eq:P1_EH_eq2}
\begin{split}
 &\min_{\bv}~~\frac{\bv^H\bQ_E\bv}{\bv^H\bQ_I\bv}\\
&\st~\bv^H\bv\leq P_T,\\
&~~~~~\bv^H\bH_E^H\bH_E\bv\geq \frac{P_{E}}{\zeta\sigma_E^2}.
\end{split}
\end{equation}

Next, we reformulate problem \eqref{eq:P1_EH_eq2} as problem \eqref{eq:P1_EH_eq_CC}. 
With variable substitution $\bv=\frac{\bu}{t}$, problem \eqref{eq:P1_EH_eq2} is equivalent to
\begin{equation}\label{eq:P1_EH_eq3}
\begin{split}
 &\min_{\bu, t}~~\frac{\bu^H\bQ_E\bu}{\bu^H\bQ_I\bu}\\
&\st~\bu^H\bu\leq P_T t^2,\\
&~~~~~\bu^H\bH_E^H\bH_E\bu\geq \frac{P_{E}}{\zeta\sigma_E^2}t^2.
\end{split}
\end{equation}
Note that the variable $t$ only appears in the two constraints of problem \eqref{eq:P1_EH_eq3}. By eliminating $t$ and combining the two constraints of problem \eqref{eq:P1_EH_eq3}, we obtain the following problem
\begin{equation}\label{eq:P1_EH_eq4}
\begin{split}
 &\min_{\bu}~~\frac{\bu^H\bQ_E\bu}{\bu^H\bQ_I\bu}\\
&\st~\bu^H\bH_E^H\bH_E\bu\geq \frac{P_{E}}{\zeta\sigma_E^2P_T}\bu^H\bu.
\end{split}
\end{equation}
It is readily known that any feasible solution $\bu$ to problem \eqref{eq:P1_EH_eq3} is feasible to problem \eqref{eq:P1_EH_eq4}. Moreover, given any feasible solution $\bar{\bu}$ to problem \eqref{eq:P1_EH_eq4}, $(\bar{\bu}, t)$ with $t=\frac{\Vert\bar{\bu}\Vert}{\sqrt{P_T}}$ is feasible to problem \eqref{eq:P1_EH_eq3}. Hence, problems \eqref{eq:P1_EH_eq3} and \eqref{eq:P1_EH_eq4} have the same feasible solution set regarding $\bu$ and thus have the same optimal solution set. Further, since scaling $\bu$ with any constant will not change the objective value while satisfying the constraint of problem \eqref{eq:P1_EH_eq4}, we can restrict $\bu^H\bQ_I\bu$ to be equal to $1$ and rewrite problem \eqref{eq:P1_EH_eq4} equivalently as \eqref{eq:P1_EH_eq_CC}.

In conclusion, the optimal solution to problem \eqref{eq:P1_EH_eq_CC}, $\bu^*$, is also an optimal solution to problem \eqref{eq:P1_EH_eq3}. By noting the relationship among problems  \eqref{eq:P1_EH_eq3}, \eqref{eq:P1_EH_eq2}, \eqref{eq:P1_EH_eq}, we conclude that $\bv^*=\sqrt{P_T}\frac{\bu^*}{\Vert\bu^*\Vert}$ is an optimal solution to problem \eqref{eq:P1_EH_eq}. This completes the proof.

\section{The Proof of Proposition \ref{prop:upperbound}}
By noting that $\bF=\bH_I^H\bH_I-\bH_E^H\bH_E\succeq0$, we have
\begin{align}
&\log\det\left(\bI+\bH_I\bX\bH_I^H\right)-\log\det\left(\bI+\bH_E\bX\bH_E^H\right)\nonumber\\
=&\log\det\left(\left(\bI+\bX\bH_E^H\bH_E+\bX\left(\bH_I^H\bH_I-\bH_E^H\bH_E\right)\right)\left(\bI+\bX\bH_E^H\bH_E\right)^{-1}\right)\nonumber\\
=&\log\det\left(\bI+\bF(\bI+\bX\bH_E^H\bH_E)^{-1}\bX\right)\nonumber\\
=&\log\det\left(\bI+\bF^{\frac{1}{2}}(\bI+\bX\bH_E^H\bH_E)^{-1}\bX\bF^{\frac{1}{2}}\right)\label{eq:obj_eq}
\end{align}
where we use the identity $\det(\bI+\bA\bB)=\det(\bI+\bB\bA)$ in the three equalities.
By replacing the objective of problem \eqref{eq:P1_EH_full} with  \eqref{eq:obj_eq} and introducing the auxiliary variable $\bY$, we rewrite problem \eqref{eq:P1_EH_full} equivalently as
\begin{equation}\label{prop2-inter-prob1}
\begin{split}
&\max_{\bX,\bY}~ \log\det\left(\bI+\bF^{\frac{1}{2}}\bY\bF^{\frac{1}{2}}\right)\\
&\st~\bY=(\bI+\bX\bH_E^H\bH_E)^{-1}\bX,\\
&~~~~~\trace(\bX)\leq P_T,\\
&~~~~~\zeta\sigma_E^2\trace(\bH_E\bX\bH_E^H)\geq P_{E},\\
&~~~~~\bX\succeq 0.
\end{split}
\end{equation}

Next, we prove that problem \eqref{prop2-inter-prob1} is equivalent to problem \eqref{eq:SCM-eq}. The proof is divided into two parts. The first part is to show that problem \eqref{prop2-inter-prob1} is equivalent to
\begin{equation}\label{prop2-intermediate-prob}
\begin{split}
&\max_{\bX,\bY} ~\log\det\left(\bI+\bF^{\frac{1}{2}}\bY\bF^{\frac{1}{2}}\right)\\
&\st~(\bI+\bX\bH_E^H\bH_E)^{-1}\bX\succeq\bY,\\
&~~~~~\trace(\bX)\leq P_T,\\
&~~~~~\zeta\sigma_E^2\trace(\bH_E\bX\bH_E^H)\geq P_{E},\\
&~~~~~\bX\succeq 0
\end{split}
\end{equation}
while the second part is to prove that problem \eqref{prop2-intermediate-prob} can be recast as problem \eqref{eq:SCM-eq}.

First, we prove the first part by showing that problems \eqref{prop2-inter-prob1} and \eqref{prop2-intermediate-prob} have the same optimal value. Let $R^*$ be the optimal value of problem \eqref{prop2-inter-prob1} and
$(\hat{\bX}, \hat{\bY})$ be an optimal solution to problem \eqref{prop2-intermediate-prob}. Since problem \eqref{prop2-intermediate-prob} is a relaxation of problem \eqref{prop2-inter-prob1}, it follows that
\begin{equation}\label{eq:ub1}
\log\det\left(\bI+\bF^{\frac{1}{2}}\hat{\bY}\bF^{\frac{1}{2}}\right)\geq R^*.
\end{equation}
On the other hand, we have
\begin{align}
&\log\det\left(\bI+\bF^{\frac{1}{2}}\hat{\bY}\bF^{\frac{1}{2}}\right)\nonumber\\
\leq& \log\det\left(\bI+\bF^{\frac{1}{2}}(\bI+\hat{\bX}\bH_E^H\bH_E)^{-1}\hat{\bX}\bF^{\frac{1}{2}}\right)\nonumber\\
\leq& R^*\label{eq:ub2}
\end{align}
where the first inequality follows from the fact that, $\det(\bI+\bA\bX_1\bA^H)\geq\det(\bI+\bA\bX_2\bA^H)$ if $\bX_1\succeq \bX_2$, and the second inequality is due to the fact that $\hat{\bX}$ is feasible to problem \eqref{prop2-inter-prob1}.
Combining \eqref{eq:ub1} and \eqref{eq:ub2}, we have $\log\det\left(\bI+\bF^{\frac{1}{2}}\hat{\bY}\bF^{\frac{1}{2}}\right)=\log\det\left(\bI+\bF^{\frac{1}{2}}(\bI+\hat{\bX}\bH_E^H\bH_E)^{-1}\hat{\bX}\bF^{\frac{1}{2}}\right)= R^*$. This implies that problems \eqref{prop2-inter-prob1} and \eqref{prop2-intermediate-prob} are equivalent.

Next we prove the second part by showing that the first constraint of problem \eqref{prop2-intermediate-prob} can be recast as a linear matrix inequality (LMI). Since $(\bI+\bH_E^H\bH_E\bX)(\bI+\bH_E^H\bH_E\bX)^{-1}=\bI$, we have
\begin{equation}\label{eq:49}
\begin{split}
\bX=(\bI+\bX\bH_E^H\bH_E)\bX-\bX(\bI+\bH_E^H\bH_E\bX)(\bI+\bH_E^H\bH_E\bX)^{-1}\bH_E^H\bH_E\bX
\end{split}
\end{equation}
By using the identity $\bX(\bI+\bH_E^H\bH_E\bX)=(\bI+\bX\bH_E^H\bH_E)\bX$ in \eqref{eq:49}, we obtain
\begin{equation}\label{eq:prop2.2-X1}
\begin{split}
\bX=(\bI+\bX\bH_E^H\bH_E)\bX-(\bI+\bX\bH_E^H\bH_E)\bX(\bI+\bH_E^H\bH_E\bX)^{-1}\bH_E^H\bH_E\bX.
\end{split}
\end{equation}
Left-multiplying $(\bI+\bX\bH_E^H\bH_E)^{-1}$ on both sides of \eqref{eq:prop2.2-X1} yields
\begin{equation}\label{eq:prop2-identity}
\begin{split}
(\bI+\bX\bH_E^H\bH_E)^{-1}\bX=\bX-\bX(\bI+\bH_E^H\bH_E\bX)^{-1}\bH_E^H\bH_E\bX.
\end{split}
\end{equation}
Using the identity $(\bI+\bA\bB)^{-1}\bA=\bA(\bI+\bB\bA)^{-1}$\cite[Sec. 3.2.4]{Mtx_book} in the RHS of Eq. \eqref{eq:prop2-identity}, we obtain
\begin{equation}\label{eq:prop2-identity2}
\begin{split}
(\bI+\bX\bH_E^H\bH_E)^{-1}\bX=\bX-\bX\bH_E^H\left(\bI+\bH_E\bX\bH_E^H\right)^{-1}\bH_E\bX
\end{split}
\end{equation}
It follows that the first constraint of problem \eqref{prop2-intermediate-prob} is equivalent to
\begin{equation}\label{eq:use_schur}
\bX-\bY\succeq\bX\bH_E^H\left(\bI+\bH_E\bX\bH_E^H\right)^{-1}\bH_E\bX.
\end{equation}
Using Schur complement\cite[Appendix A.5.5]{cvx_book}, we  can write \eqref{eq:use_schur} equivalently as the following LMI
$$\left[\begin{array}{cc}\bX-\bY&\bX\bH_E^H\\
\bH_E\bX&\bI+\bH_E\bX\bH_E^H
\end{array}\right]\succeq0.$$
Therefore, by replacing the first constraint in problem \eqref{prop2-intermediate-prob} with the above LMI and noting that the resulting problem is convex, we complete the proof.

\section{The Proof of Proposition \ref{prop:stationary}}
For ease of exposition, we denote problem \eqref{eq:subpV-linearization} by $\mP(\tV, \bU, \bW_I, \bW_E)$, its solution set by $\bbS(\tV, \bU, \bW_I, \bW_E)$, and its constraint set by $\mC_{\leq}(\tV)$. Let $\{\bV^k, \bU^k, \bW_I^k, \bW_E^k\}$ denote the iterates generated by the IBCD algorithm in TABLE II, where $\bU^k$, $\bW_I^k$, and $\bW_E^k$ are obtained via Steps 4-6 with $\tV=\bV^k$, and $\bV^{k}$ is obtained (via Step 7) by solving problem  $\mP(\bV^{k-1}, \bU^{k-1}, \bW_I^{k-1}, \bW_E^{k-1})$. Denote by $f(\bV, \bU, \bW_I, \bW_E)$  the objective function of problem \eqref{eq:P1_EH_WMMSE}. Moreover, define $g(\bV)\triangleq\trace(\bV^H\bH_E^H\bH_E\bV)$ and
\begin{equation}
\begin{split}
\bg(\bV, \tV)\triangleq\trace(\tilde{\bV}^H\bH_E^H\bH_E\tilde{\bV})+\trace(\tilde{\bV}^H\bH_E^H\bH_E(\bV-\tilde{\bV}))+\trace((\bV-\tilde{\bV})^H\bH_E^H\bH_E\tilde{\bV}).
\end{split}
\end{equation}
It follows that $\bg(\bV, \bV)=g(\bV)$. Moreover, it can be easily verified that $f(\bV^{k}, \bU^k, \bW_I^k, \bW_E^k)=C(\bV^k)$ by noting $\bW_I^k=\mE(\bU^k, \bV^k)^{-1}$.
In the following, we complete the proof through four steps.

\emph{\underline{In the first step}}, we show that each $\bV^{k}$ for $k=1,2,\ldots$ is feasible to problem \eqref{eq:P1_EH_WMMSE}. It suffices to show that $\bV^{k+1}$ is feasible to problem \eqref{eq:P1_EH_WMMSE} if $\bV^k$ is. Assume that $\bV^{k}$ is feasible to problem \eqref{eq:P1_EH_WMMSE}. Thus, we have $\bg(\bV^{k}, \bV^k)=g(\bV^{k})\geq \frac{P_E}{\zeta\sigma_E^2}$ and $\trace\left(\bV^k\left(\bV^k\right)^H\right)\leq P_T$. It follows that there must exist $\bV^{k+1}$ that is feasible to problem $\mP(\bV^{k}, \bU^k, \bW_I^k, \bW_E^k)$. Thus we have $\bV^{k+1}\in \mC_{\leq}(\bV^{k})$, that is, $\bV^{k+1}$ is such that $\trace\left(\bV^{k+1}(\bV^{k+1})^H\right)\leq P_T$ and $\bg(\bV^{k+1}, \bV^k)\geq \frac{P_E}{\zeta\sigma_E^2}$.
Furthermore, since $g(\bV)$ is a convex function in $\bV$, we have $g(\bV)\geq \bg(\bV, \tV)$ for any $\bV$ and $\tV$\cite{cvx_book}. It follows that
\begin{equation}\label{eq:key_ineq_conv}
g(\bV^{k+1})\geq \bg(\bV^{k+1}, \bV^k)\geq \frac{P_E}{\zeta\sigma_E^2}
\end{equation}
which together with the fact $\trace\left(\bV^{k+1}(\bV^{k+1})^H\right)\leq P_T$ implies that $\bV^{k+1}$ is feasible to problem \eqref{eq:P1_EH_WMMSE}. Thus the first step is finished.

\emph{\underline{In the second step}}, we show that the objective value sequence $\{C(\bV^k)\}$ has monotonic convergence. We have
\begin{equation}\label{eq:obj_descend}
\begin{split}
C(\bV^{k+1})&=f(\bV^{k+1}, \bU^{k+1}, \bW_I^{k+1}, \bW_E^{k+1})\\
&\geq f(\bV^{k+1}, \bU^{k+1}, \bW_I^k, \bW_E^k)\\
&\geq f(\bV^{k+1}, \bU^k, \bW_I^k, \bW_E^k)\\
&\geq f(\bV^{k}, \bU^k, \bW_I^k, \bW_E^k)=C(\bV^k)
\end{split}
\end{equation}
 where the first inequality is due to Steps 5 and 6 (i.e., Sub-iteration 2), the second inequality is due to Step 4 (i.e., Sub-iteration 1), and the third inequality is due to Step 7 (i.e., Sub-iteration 3) and that $\bV^k$ is a feasible solution to problem $\mP(\bV^{k}, \bU^k, \bW_I^k, \bW_E^k)$. Since $C(\bV^k)$ is upper bounded due to the compactness of $\{\bV^k\}$ and the continuity of $C(\bV)$, the inequality \eqref{eq:obj_descend} leads to the monotonic convergence of $\{C(\bV^k)\}$. Thus the second step is finished.

\emph{\underline{In the third step}}, we prove that any limit point $\left(\bV^{*}, \!\bU^*, \!\bW_I^*, \!\bW_E^*\right)$ of the iterates $\{\bV^{k}, \bU^k, \bW_I^k, $\\$\bW_E^k\}$ is a KKT point of problem \eqref{eq:P1_EH_WMMSE}. The proof is by first showing $\bV^{*}\in \bbS(\bV^*, \bU^*, \bW_I^*, \bW_E^*)$ and then arguing that $\left(\bV^{*}, \bU^*, \bW_I^*, \bW_E^*\right)$ satisfy the KKT condition of problem \eqref{eq:P1_EH_WMMSE}.

We first prove $\bV^{*}\in \bbS(\bV^*, \bU^*, \bW_I^*, \bW_E^*)$.
Since $\left(\bV^{*}, \bU^*, \bW_I^*, \bW_E^*\right)$ is a limit point of $\{\bV^{k}, \bU^k, \bW_I^k, \bW_E^k\}$, there must exist a convergent subsequence $\{\bV^{k_j}, \bU^{k_j}, \bW_I^{k_j}, \bW_E^{k_j}\}$ such that $\lim_{j\rightarrow \infty}\bV^{k_j} =\bV^*$. Due to the compactness of the constraint set of problem \eqref{eq:P1_EH_WMMSE}, by restricting to a subsequence, we can assume that $\{\bV^{k_j+1}\}$ converges to a limit point $\bV^{**}$.


Define $\mC_{<}(\tV)\triangleq
\left\{\bV~|~\trace(\bV\bV^H)\leq P_T, \bg(\bV, \tV)> \frac{P_E}{\zeta\sigma_E^2}\right\}$. It follows that $\mC_{<}(\tV)\subset \mC_{\leq}(\tV)$ for any $\tV$. Let us consider the set $\mC_{<}(\bV^*)$. Since $\bg(\bV, \tV)$ is continuous in $\tV$ and $\lim_{j\rightarrow \infty} \bV^{k_j}=\bV^*$, there must exist, for any fixed $\bV\in \mC_{<}(\bV^*)$,  an integer $I_{\bV}$ such that
$$\bg(\bV, \bV^{k_j})> \frac{P_E}{\zeta\sigma_E^2}, ~\forall j\geq I_{\bV}.$$
This implies that, there exists a sufficiently large $I$ such that $$\mC_{<}(\bV^*)\subseteq\mC_{<}(\bV^{k_j})\subset\mC_{\leq}(\bV^{k_j}), ~\forall j>I.$$  Since
$\bV^{k_j+1}\in \bbS(\bV^{k_j}, \bU^{k_j}, \bW_I^{k_j}, \bW_E^{k_j})$, we have
\begin{equation}\label{eq:V_k_jopt}
\begin{split}
f(\bV, \bU^{k_j}, \bW_I^{k_j}, \bW_E^{k_j})\leq f(\bV^{k_j+1}, \bU^{k_j}, \bW_I^{k_j}, \bW_E^{k_j}), ~\forall \bV\in\mC_{<}(\bV^*)\subset\mC_{\leq}(\bV^{k_j}).
\end{split}
\end{equation}
Moreover, since $f(\cdot)$ is a continuous function, we have by letting $j\rightarrow \infty$ in \eqref{eq:V_k_jopt}
\begin{equation}
\begin{split}
f(\bV, \bU^{*}, \bW_I^{*}, \bW_E^{*})\leq f(\bV^{**}, \bU^{*}, \bW_I^{*}, \bW_E^{*}),~\forall \bV\in\mC_{<}(\bV^*).
\end{split}
\end{equation}
It follows from the continuity of $\bg(\bV, \tV)$ that
\begin{equation}\label{eq:Vopt_prop}
\begin{split}
f(\bV, \bU^{*}, \bW_I^{*}, \bW_E^{*})\leq f(\bV^{**}, \bU^{*}, \bW_I^{*}, \bW_E^{*}),~\forall \bV\in\mC_{\leq}(\bV^*).
\end{split}
\end{equation}
On the other hand, \eqref{eq:obj_descend} implies
\begin{equation}\label{eq:V*=V**}
f(\bV^*, \bU^{*}, \bW_I^{*}, \bW_E^{*})= f(\bV^{**}, \bU^{*}, \bW_I^{*}, \bW_E^{*}).
\end{equation}
Moreover, since $\bV^{k_j}$ is feasible to problem \eqref{eq:P1_EH_WMMSE} and $\bg(\bV^{k_j}, \bV^{k_j})=g(\bV^{k_j})$, we have $\bV^{k_j}\in\mC_{\leq}(\bV^{k_j})$. It follows that $\bV^*\in\mC_{\leq}(\bV^*)$.
Combining this with \eqref{eq:Vopt_prop} and  \eqref{eq:V*=V**}, we obtain $\bV^{*}\in \bbS(\bV^*, \bU^*, \bW_I^*, \bW_E^*)$.

Then we show that $(\bV^*, \bU^*, \bW_I^*, \bW_E^*)$ is a KKT point of problem \eqref{eq:P1_EH_WMMSE}. Since Slater's condition holds for problem $\mP(\bV^*, \bU^*, \bW_I^*, \bW_E^*)$ and $\bV^{*}\in \bbS(\bV^*, \bU^*, \bW_I^*, \bW_E^*)$, there exists optimal Lagrange multipliers $\lambda^*$ and $\mu^*$, together with $\bV^*$, satisfying the KKT conditions\cite{cvx_book} of problem $\mP(\bV^*, \bU^*, \bW_I^*, \bW_E^*)$, i.e.,
\begin{align}
&\left(\bH_I^H\bU^*\bW_I^*\left(\bU^*\right)^H\bH_I+\bH_E^H\bW_E^*\bH_E+\lambda^*\bI\right)\bV^*-\bH_I^H\bU^*\bW_I^*-\mu^*\bH_E^H\bH_E\bV^*=0,\label{eq:subpV-KKT1}\\
&\lambda^*\left(\trace\left(\bV^*\left(\bV^*\right)^H\right)-P_T\right)=0,\label{eq:subpV-KKT2}\\
&\mu^*\left(\trace\left(\left(\bV^*\right)^H\bH_E^H\bH_E\bV^*\right)-\frac{P_E}{\zeta\sigma_E^2}\right)=0,\label{eq:subpV-KKT3}\\
&\trace(\bV^*\left(\bV^*\right)^H)\leq P_T,\label{eq:subpV-KKT4}\\
&\trace\left(\left(\bV^*\right)^H\bH_E^H\bH_E\bV^*\right)\geq\frac{P_E}{\zeta\sigma_E^2},\label{eq:subpV-KKT5}\\
&\lambda^*, \mu^*\geq 0\label{eq:subpV-KKT6}
\end{align}
where \eqref{eq:subpV-KKT1} is the first-order necessary optimality condition, \eqref{eq:subpV-KKT2} and \eqref{eq:subpV-KKT3} are the complementarity conditions,  \eqref{eq:subpV-KKT4} and \eqref{eq:subpV-KKT5} are the primal feasibility conditions, and \eqref{eq:subpV-KKT6} is the dual feasibility condition. Note that we have used the fact $g(\bV^*)=\bg(\bV^*, \bV^*)$ in \eqref{eq:subpV-KKT3} and \eqref{eq:subpV-KKT5}.

On the other hand, by the continuity we have
\begin{align}
&\bU^*=\left(\bI+\bH_I\bV^*\left(\bV^*\right)^H\bH_I^H\right)^{-1}\bH_I\bV^* \label{eq:subpV-KKT7},\\
&\bW_I^* = \bI+\bH_I\bV^*\left(\bV^*\right)^H\bH_I^H\label{eq:subpV-KKT8},\\
&\bW_E^* = \left(\bI+\bH_E\bV^*\left(\bV^*\right)^H\bH_E^H\right)^{-1}\label{eq:subpV-KKT9}.
\end{align}
Eqs. \eqref{eq:subpV-KKT1}-\eqref{eq:subpV-KKT9} imply that $(\bV^*, \bU^*, \bW_I^*, \bW_E^*)$  is a KKT point of problem \eqref{eq:P1_EH_WMMSE}. Thus the third step is finished.

\emph{\underline{In the last step}}, we prove that $\bV^*$ is a KKT point of problem \eqref{eq:P1_EH} by reducing Eqs. \eqref{eq:subpV-KKT1}--\eqref{eq:subpV-KKT9} to the KKT conditions of problem \eqref{eq:P1_EH}. Let us consider the term $\bH_I^H\bU^*\bW_I^*(\bU^*)^H\bH_I\bV^*$ in \eqref{eq:subpV-KKT1}. According to Fact 2 in Lemma \ref{lem:key_idea_WMMSE}, we have $(\bW_I^*)^{-1}=\bI-\bU^*\bH_I\bV^*$. It follows that
\begin{equation}\label{eq:term45}
\begin{split}
\bH_I^H\bU^*\bW_I^*(\bU^*)^H\bH_I\bV^*&=\bH_I^H\bU^*\bW_I^*\left(\bI-(\bW_I^*)^{-1}\right)\\
&=\bH_I^H\bU^*\bW_I^*-\bH_I^H\bU^*.
\end{split}
\end{equation}
Substituting \eqref{eq:term45} into \eqref{eq:subpV-KKT1}, we simplify \eqref{eq:subpV-KKT1} to
\begin{equation}\label{eq:P1-KKT}
-\bH_I^H\bU^*+\left(\bH_E^H\bW_E^*\bH_E+\lambda^*\bI\right)\bV^*-\mu^*\bH_E^H\bH_E\bV^*=0
\end{equation}
Further, plugging \eqref{eq:subpV-KKT7} and \eqref{eq:subpV-KKT9} into \eqref{eq:P1-KKT}, we have
\begin{equation}\label{eq:P1-KKT1}
\begin{split}
&\left(-\bH_I^H\left(\bI+\bH_I\bV^*\left(\bV^*\right)^H\bH_I^H\right)^{-1}\bH_I\right.\\
&~~~~~~~~~\left.+\bH_E^H\left(\bI+\bH_E\bV^*\left(\bV^*\right)^H\bH_E^H\right)^{-1}\bH_E+\lambda^*\bI-\mu^*\bH_E^H\bH_E\right)\bV^*=0
\end{split}
\end{equation}
Eqs. \eqref{eq:P1-KKT1} and \eqref{eq:subpV-KKT2}-\eqref{eq:subpV-KKT6} imply that $\bV^*$ is a KKT point of problem \eqref{eq:P1_EH}. This completes the proof.

\ifCLASSOPTIONcaptionsoff
  \newpage
\fi


\end{document}